\documentclass{ws-procs9x6}

\newcommand{\Z}{Z\!\!\!Z}

\begin{document}

\title{When is the deconfinement phase transition universal?}

\author{K. HOLLAND}

\address{Department of Physics, 
	 University of California at San Diego, \\
	 La Jolla CA 92093-0319, USA}

\author{M. PEPE AND U. -J. WIESE\footnote{on leave from \uppercase{MIT}.}}

\address{Institute for Theoretical Physics, 
         University of Bern, \\
	 Sidlerstrasse 5, CH-3012 Bern, Switzerland}


\maketitle

\abstracts{
Pure Yang-Mills theory has a finite-temperature phase transition, separating 
the confined and deconfined bulk phases. Svetitsky and Yaffe conjectured that 
if this phase transition is of second order, it belongs to the universality 
class of transitions for particular scalar field theories in one lower
dimension. We examine Yang-Mills theory with the symplectic gauge groups 
$Sp(N)$. We find new evidence supporting the Svetitsky-Yaffe conjecture 
and make our own conjecture as to which gauge theories have a universal 
second order deconfinement phase transition.}

\section{INTRODUCTION}

Yang-Mills theories with gauge group $G$ have a finite-temperature phase 
transition, separating the confined phase of colorless glueballs from the 
deconfined gluon plasma phase\cite{Holland:2000uj}. The transition 
is signalled by the spontaneous breaking of a global 
symmetry\cite{'tHooft:1977hy} related to $H$, the center of $G$. The 
Polyakov loop is the order parameter for the transition\cite{Polyakov:vu}, 
transforming as $\Phi(\vec{x})' = z \Phi(\vec{x}), z \in H$ under the global 
center transformation. Its expectation value is $\langle \Phi \rangle = 
\exp(-\beta F_q)$, where $F_q$ is the free energy of a static quark 
in the gluon background, $\beta = 1/T$ is the time extent and
$T$ is the temperature. In the confined phase, there are no isolated quarks
and $F_q \rightarrow \infty$ in the infinite volume limit, giving 
$\langle \Phi \rangle = 0$. In the deconfined phase, $F_q$ is finite and
$\langle \Phi \rangle \ne 0$, spontaneously breaking the global center
symmetry.

Svetitsky and Yaffe conjectured that if Yang-Mills theory with gauge group
 $G$ has a second order deconfinement transition, with the correlation length 
$\xi \rightarrow \infty$, it should have universal properties belonging to the 
universality class of $H$-symmetric scalar field theories in one lower
dimension\cite{Svetitsky:1982gs}. When $\xi \gg \beta$, the system is no 
longer sensitive to the finite time extent and dimensional reduction occurs. 
The universality class is determined by quantities such as the critical 
exponents, e.g.~how the correlation length diverges as the critical 
temperature is approached, $\xi \propto (T - T_{\rm c})^{-\nu}$. Note, 
however, that the conjecture does {\it not} state that the deconfinement 
transition {\it must} be second order.

Most studies have looked at 4-d and 3-d $SU(N)$ Yang-Mills theories. The 
center of $SU(N)$ is $\Z(N)$, the $N$ roots of unity. We first consider 4-d 
theories. With $SU(2)$ as the gauge group, the theory has a second order 
deconfinement transition\cite{McLerran:1980pk}, belonging to the universality 
class of 3-d $\Z(2)$-symmetric scalar field theory, i.e.~the Ising 
universality class\cite{Engels:1989fz}. This fully supports the 
Svetitsky-Yaffe conjecture. For $SU(3)$, the transition is weakly first order
with a large but finite correlation length $\xi$ and no universal 
properties\cite{Celik:1983wz}. Studies have shown that for $SU(N)$ 
theories with $N=4,6,8$, the deconfinement transitions are first order,
with the strength increasing for larger $N$, again without any universal
 properties\cite{Lucini:2003zr}. Interestingly, 3-d $\Z(N)$-symmetric scalar 
field theory for $N \geq 5$ is in the universality class of the 3-d 
$U(1)$-symmetric XY model\cite{Hove03}. The gauge theories simply don't make 
use of this universality class. There is a richer structure in 3 dimensions. 
For $SU(N)$ with $N=2,3,4$, the deconfinement transitions are second order,
belonging to the 2-d $\Z(N)$-symmetric universality classes with $N=2,3,4$ 
respectively, again supporting the Svetitsky-Yaffe 
conjecture\cite{Engels:1996dz}.

\section{$Sp(N)$ GAUGE THEORY}

\begin{figure}[thb]
\centerline{
\epsfxsize=6.5cm\epsfbox{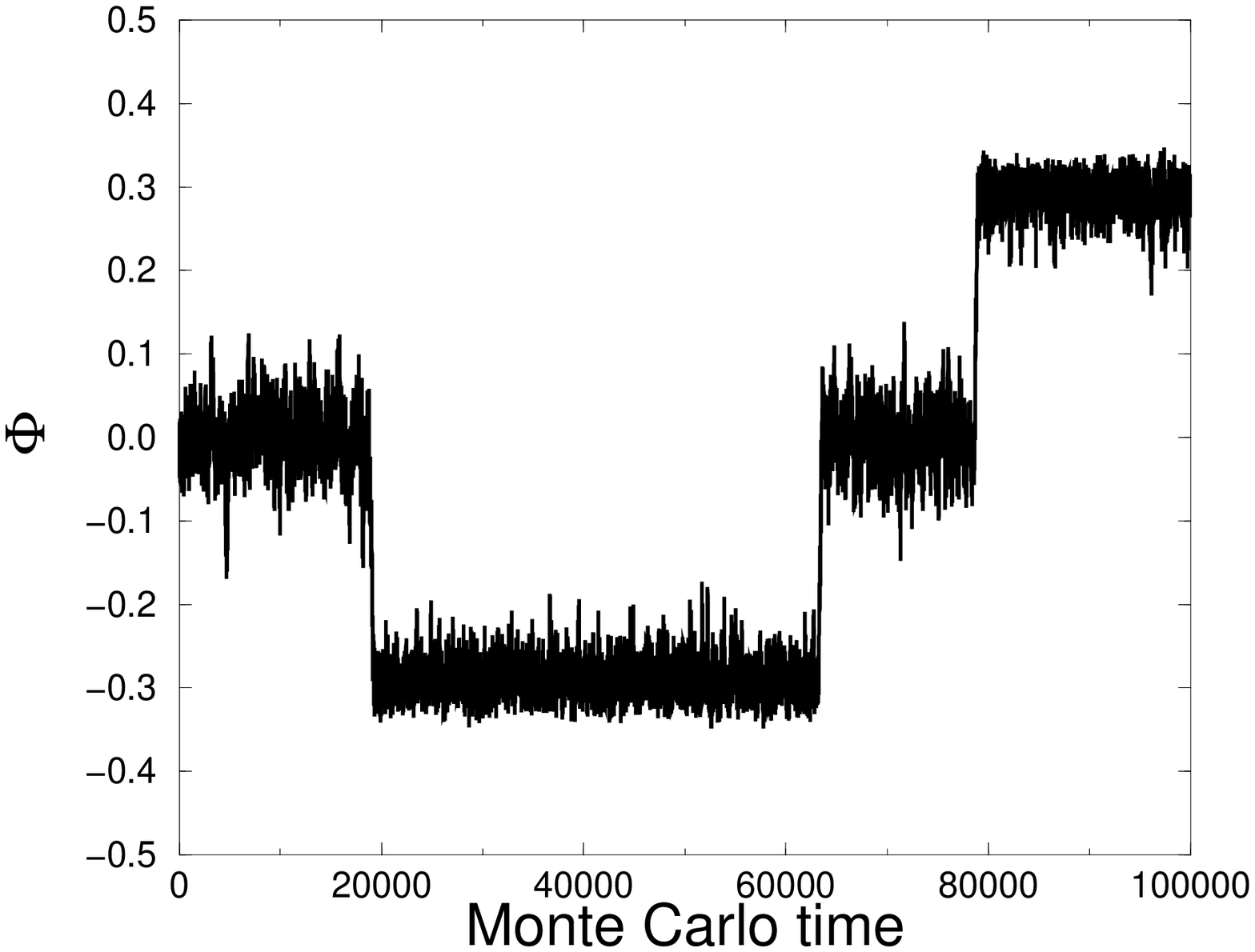}
\epsfxsize=6.5cm\epsfbox{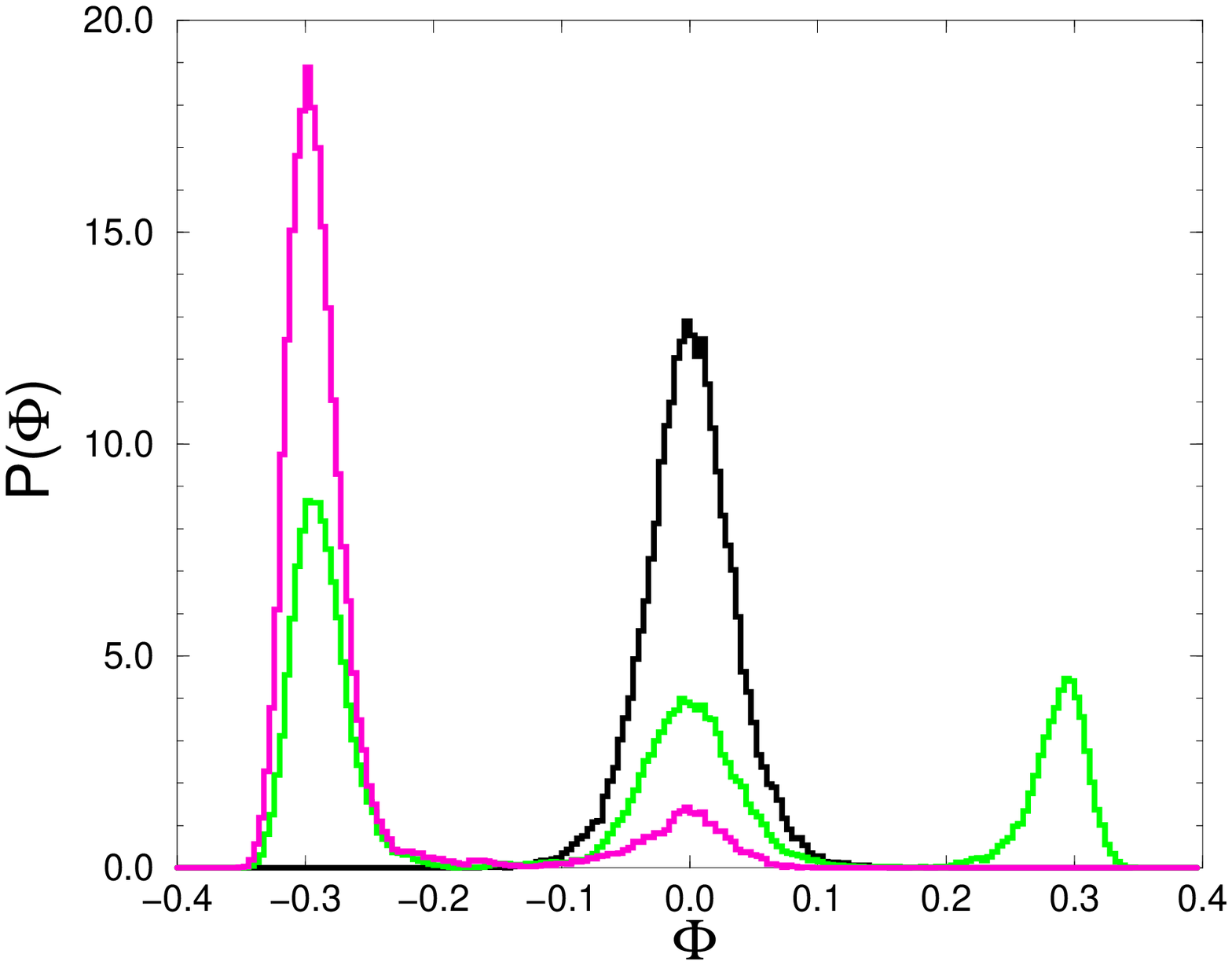}
}   
\caption{(a) Tunneling between coexisting confined and deconfined phases
	at $T_{\rm c}$ and (b) probability distributions of $\Phi$ for 
	temperatures close to $T_{\rm c}$.}
\label{fig:polyakov}
\end{figure}

We look at Yang-Mills theories with the symplectic groups $Sp(N)$. These 
groups have the property that $Sp(N) \subset SU(2N)$ and $U \in Sp(N)$ 
satisfies the constraint 
\begin{equation}
U^\ast = J U J^\dagger, \hspace{0.5cm} 
J= \left( \begin{array}{cc}0&1\!\!1\\-1\!\!1&0 \end{array} \right)
\hspace{0.5cm} \Rightarrow \hspace{0.5cm} 
U=\left( \begin{array}{cc}W&X\\-X^\ast&W^\ast \end{array} \right),
\end{equation}
where $W$ and $X$ are complex $N \times N$ matrices. Since $U$ and $U^\ast$
are related by a unitary transformation $J \in Sp(N)$, the $2N$-dimensional
fundamental representation of $Sp(N)$ is pseudo-real and charge conjugation
is a global gauge transformation. These properties are familiar from $SU(2)$. 
Writing $U = \exp(iH)$, the Hermitean matrix $H$ satisfies
\begin{equation}
H^\ast = -J H J^\dagger = J H J \hspace{0.5cm}  \Rightarrow \hspace{0.5cm} 
H=\left( \begin{array}{cc}A&B\\B^\ast&-A^\ast \end{array} \right),
\end{equation}
where the $N \times N$ complex matrices $A$ and $B$ satisfy $A=A^\dagger$ and
$B=B^T$. The $N^2$ and $N(N+1)$ degrees of freedom of $A$ and $B$ respectively
mean that $Sp(N)$ has $N^2+N(N+1)=(2N+1)N$ generators. $Sp(N)$ has
rank $N$. There are the special equivalent cases $Sp(1) = SU(2)$ and 
$Sp(2) \simeq SO(5)$. Most interestingly, unlike $SU(N)$, the center of 
$Sp(N)$ is $\Z(2)$ for all $N$. This allows us to disentangle the size of the 
group from the center and see what effect this has on the deconfinement 
transition. According to the Svetitsky-Yaffe conjecture, a second order 
transition should belong to the Ising universality class.

The lattice formulation of $Sp(N)$ Yang-Mills theory is straightforward.
We use the Wilson action in our simulations. As $SU(2) \subset Sp(N)$, we
can update the $Sp(N)$ gauge links using the standard 
heatbath\cite{Creutz:zw} and overrelaxation\cite{Cabibbo:zn} algorithms to 
update the various $SU(2)$ subgroups, in the same way as done for $SU(N)$ 
gauge theory. We find that there is no bulk phase transition between strong 
and weak coupling. Further details will be presented in a forthcoming 
paper\cite{Holland03}.

\begin{figure}[thb]
\centerline{
\epsfxsize=6.5cm\epsfbox{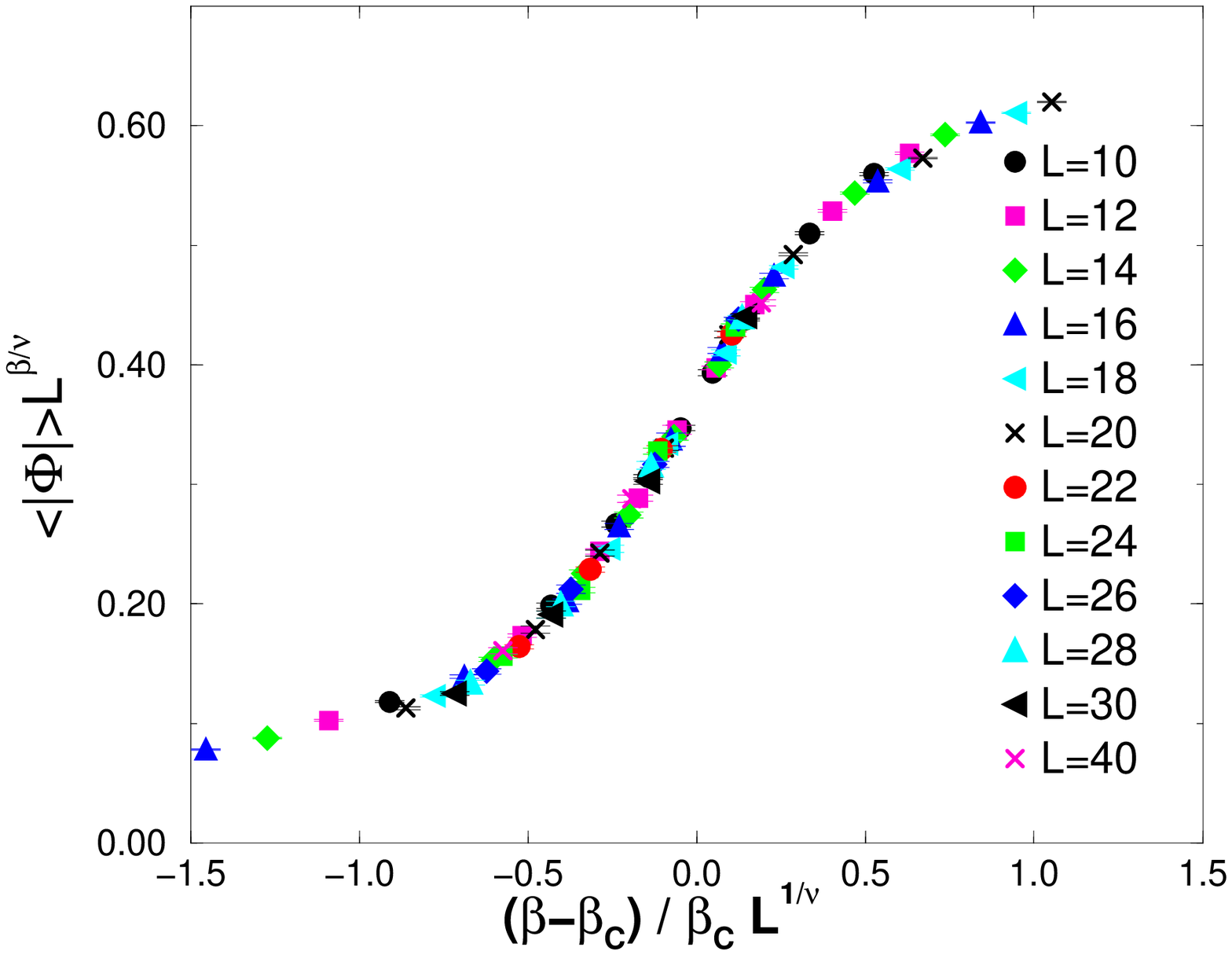}
\epsfxsize=6.5cm\epsfbox{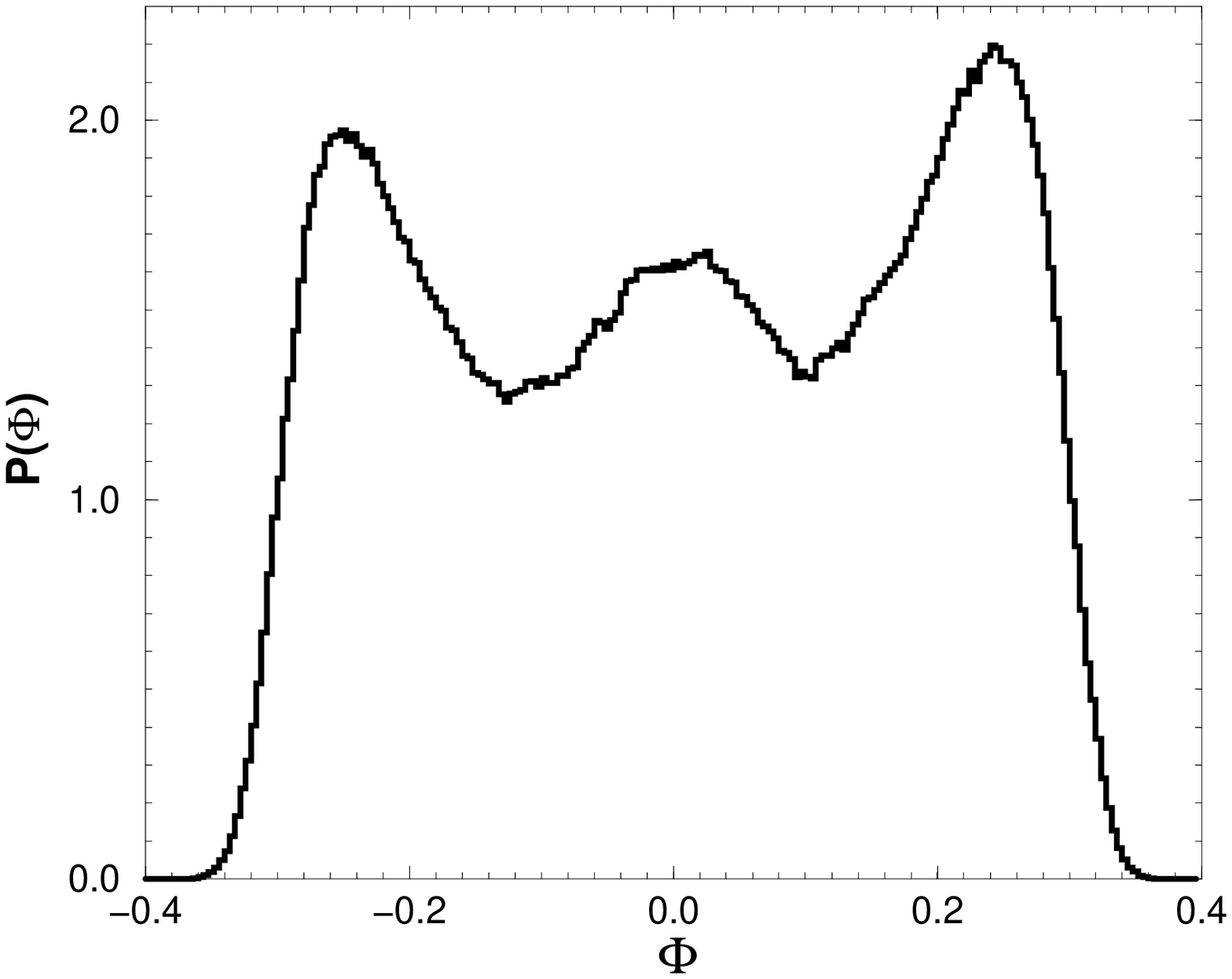}
}   
\caption{(a) The universal function $\langle |\Phi| \rangle L^{\beta/\nu}$
	for 3-d $Sp(2)$ gauge theory and 
	(b) coexisting confined and deconfined phases for 
	3-d $Sp(3)$ gauge theory.}
\label{fig:3dim}
\end{figure}

We first consider 4-d $Sp(2)$ gauge theory, where one might expect the
deconfinement transition to be second order, just as for $SU(2)$. However,
we find the transition to be first order, even with $\Z(2)$ as the center. In 
Fig.~\ref{fig:polyakov}, we plot tunneling  between coexisting confined 
($\Phi = 0$) and deconfined ($\Phi \ne 0$) phases at the critical 
temperature, as well as the probability distributions of $\Phi$ close to 
$T_{\rm c}$. Coexistence of the phases is a clear signal for a first order 
transition. Measurements of the Polyakov loop susceptibility, the specific 
and latent heats also show a clear first order transition. Similarly, we find 
that 4-d $Sp(3)$ Yang-Mills theory also has a first order deconfinement 
transition. Going from $Sp(1) = SU(2)$ to $Sp(2)$, the phase transition 
changes from second to first order, even though the center of the 
group is the same, indicating that the size of the group is more important.

In 3-d, we find a richer structure. Exactly like $Sp(1)$, we find that 
$Sp(2)$ Yang-Mills theory has a second order deconfinement transition.
Measurements for various temperatures and volumes can be mapped onto one 
universal curve when rescaled using the critical exponents of the 2-d Ising 
universality class, as shown in Fig.~\ref{fig:3dim}(a). This is new evidence 
supporting the Svetitsky-Yaffe conjecture. However, for $Sp(3)$, the 
transition is weakly first order, as we see in Fig.~\ref{fig:3dim}(b), where 
one has to go to large volumes to distinguish the coexisting phases. 
Again, we find that the transition switches from second to first order as we 
increase the size of the group, even though the center is unchanged.

\section{SUMMARY}
From our work and other studies, we conjecture that only for $Sp(1) =
SU(2) \simeq SO(3)$ is there a universal second order deconfinement phase
transition in 4 dimensions. In 3-d, we find that $Sp(2)$ gauge theory has
a second order deconfinement transition belonging to the 2-d Ising universality
class, which is new evidence supporting the Svetitsky-Yaffe conjecture. We 
expect that there are no other second order transitions with other gauge 
groups\cite{Holland03,Holland:2003jy}. In both 4-d and 3-d $Sp(N)$ gauge 
theory, we find that the deconfinement transition changes from second to 
first order as we increase $N$, even though the center of the group is 
always $\Z(2)$. The order of the transition seems to be dictated by the size 
of the group, not the center. This is natural, as the number of glueballs in 
the confined phase is group-independent, whereas the number of deconfined 
gluons increases with the group size, leading to a larger mismatch in the 
number of degrees of freedom at the critical temperature as $N$ increases.

\section*{ACKNOWLEDGMENTS}

This work has been supported under grant DOE-FG03-97ER40546,
by the Schweizerischer Nationalfond and the European Community's Human 
Potential Program HPRN-CT-2000-00145 Hadrons/Lattice QCD.

\end{document}